\documentclass{edm_article}
\usepackage{url}
\usepackage{multirow}
\usepackage[table]{xcolor}  
\usepackage{caption}  
\usepackage{wrapfig}
\newcounter{copyrightbox}
\usepackage{newfloat}
\usepackage{booktabs}
\usepackage{siunitx}   

\newcommand{\aysa}[1]{\textbf{\textcolor{blue}{}}}

\begin{document}

\title{Evaluating the Quality of Code Comments Generated by Large Language Models for Novice Programmers}

\numberofauthors{4}
\author{
\alignauthor
Aysa Xuemo Fan\\
       \affaddr{University of Illinois at Urbana-Champaign}\\
       \email{xuemof2@illinois.edu}
\alignauthor
Arun Balajiee Lekshmi Narayanan\\
       \affaddr{University of Pittsburgh}\\
       \email{arl122@pitt.edu}
\alignauthor 
Mohammad Hassany\\
       \affaddr{University of Pittsburgh}\\
       \affaddr{Carnegie Mellon University	}\\
       \email{moh70@pitt.edu}
\and  
\alignauthor
Jiaze Ke\\
       \affaddr{Carnegie Mellon University}\\
       \email{jiazek@andrew.cmu.edu	}
}

\maketitle

\begin{abstract}
Large Language Models (LLMs) show promise in generating code comments for novice programmers, but their educational effectiveness remains under-evaluated. This study assesses the instructional quality of code comments produced by GPT-4, GPT-3.5-Turbo, and Llama2, compared to expert-developed comments, focusing on their suitability for novices. Analyzing a dataset of ``easy'' level Java solutions from LeetCode, we find that GPT-4 exhibits comparable quality to expert comments in aspects critical for beginners, such as clarity, beginner-friendliness, concept elucidation, and step-by-step guidance. GPT-4 outperforms Llama2 in discussing complexity (chi-square = 11.40, p = 0.001) and is perceived as significantly more supportive for beginners than GPT-3.5 and Llama2 with Mann-Whitney U-statistics = 300.5 and 322.5, p = 0.0017 and 0.0003). This study highlights the potential of LLMs for generating code comments tailored to novice programmers.
\end{abstract}


\section{Introduction}
As the field of artificial intelligence (AI) evolves, the integration of Large Language Models (LLMs) into educational contexts is gaining interest \cite{yan2023practical, baierl2023applications, meyer2023chatgpt}, particularly in the realm of computer science education for novice programmers \cite{leinonen2023using, macneil2022automatically}. LLMs show promise in generating code comments, a valuable resource for novice programmers grappling with complex programming concepts \cite{macneil2022automatically, leinonen2023comparing, macneil2022generating}. Recent works by Cui et al. \cite{Cui2022CodeExpEC}, Codex\footnote{https://openai.com/blog/openai-codex}, and Madaan et al. \cite{Madaan2022LanguageMO} explore the development of code comments using docstrings as training material for AI models, highlighting the potential of LLMs in supporting novice programmers.

While studies by Cui et al. \cite{cui2022codeexp} and Madaan et al. \cite{madaan2022language} demonstrate the feasibility of using LLMs to generate code comments, the educational meaningfulness and effectiveness of these comments for novice programmers have been under-evaluated. Leinonen et al. \cite{leinonen2023comparing} emphasize the need for further research on the quality and instructional value of LLM-generated comments for novice programmers compared to human-crafted ones, stressing the importance of assessing their impact on learning outcomes and student engagement. 
While MacNeil et al. [9] emphasize the importance of incorporating end-user insights, our study focuses on expert evaluations of LLM-generated comments for novice programmers. Future research will aim to include direct feedback from novice users to further validate these findings.
Understanding the performances, strengths and weaknesses of LLM-generated comments and how they compare to expert-developed comments can contribute to refining educational approaches in programming, particularly for novice programmers. While \cite{macneil2022generating} suggest the importance of user studies with novice programmers, our current study focuses on expert evaluation of LLM-generated comments. Future work will incorporate direct feedback from novice programmers.

To address this, our study focuses on evaluating the instructional quality of code comments produced by LLMs for novice programmers. 
We developed a codebook for this analysis, using predefined criteria to systematically evaluate the instructional quality of the comments for novice programmers.
The study utilizes a dataset from LeetCode\footnote{\url{https://leetcode.com}}, a well-known competitive programming platform. The dataset includes Java programming solutions, which serve as input for the LLMs to generate comments tailored to novice programmers. The inline documentation provided by experienced programmers in the LeetCode solutions is used as a benchmark for comparing the quality of LLM-generated comments for novice programmers.

We aim to answer three research questions:

1. How do LLM-generated code comments compare to expert-developed comments based on evaluation criteria designed to assess their instructional quality and effectiveness?

2. What are the strengths and weaknesses of LLM-generated code comments in terms of criteria?

3. Based on established evaluation criteria, which source of code comments aligns better with the instructional needs of novice programmers?

By focusing on the evaluation metrics and aspects, we aim to provide insights that can inform the development and refinement of LLM-based tools for generating instructional content in computer science education, rather than limiting the conclusions to specific LLMs.

Our study contributes to the field of computer science education and AI-driven educational resources for novice programmers in several ways. First, we propose a comprehensive codebook designed specifically for assessing the instructional quality of code comments generated by LLMs for novice programmers. This codebook can serve as a valuable tool for educators and researchers to evaluate the effectiveness of AI-generated comments in programming education for novice programmers. Second, our study identifies the strengths and weaknesses of LLM-generated comments across various aspects of instructional quality relevant to novice programmers, such as clarity, beginner-friendliness, concept elucidation, and step-by-step guidance. These insights can inform the development and refinement of LLM-based tools for generating instructional content tailored to novice programmers in computer science education. Finally, by comparing the performance of different state-of-the-art LLMs in generating code comments for novice programmers, our study provides guidance for educators and researchers interested in leveraging LLMs for educational purposes. The findings can help in selecting the most suitable LLMs for specific educational contexts and identifying areas where further improvements are needed to better support novice programmers.

\section{Methods}
\subsection{Dataset Selection and Prompting}
Our study's methodology is specifically designed to evaluate the effectiveness of LLM-generated code comments for novice programmers. To this end, we have strategically selected the LeetCode Java Solution dataset\footnote{\url{https://github.com/cheehwatang/leetcode-java}} which comprises solutions to 30 ``easy'' level problems. This choice is intentional; the ``easy'' level categorization aligns with the introductory level of computer science education and offers a range of concepts that are suitable and accessible for beginner-level programmers. By focusing on simpler coding problems, our study aims to provide insights into the helpfulness of LLM-based comments in fostering understanding and learning in an intro-to-cs context. This approach ensures that the complexity of the problems does not overshadow the educational potential of the comments provided by the language models.
The LLMs were prompted to explain existing code solutions, simulating a scenario where a novice programmer seeks explanation of a correct solution and it explicitely asked for the complexity, overview and step-by-step comments.

\subsection{Model Selection}
We adapted the prompt structure, and also followed the prompt engineering methodology from \cite{li2023explaining}, structuring prompts to ensure compatibility with various LLMs and alignment with expert-level commentary. Three advanced LLMs were chosen: GPT-4 \cite{openai2023gpt4}, Llama2 \cite{touvron2023llama}, and GPT-3.5-Turbo \cite{ye2023comprehensive}. These models were selected for their demonstrated performance in similar tasks and their advanced language understanding abilities \cite{xu2022systematic, ye2023comprehensive}. The full prompts is in the Appendix. In the 2 phases of evaluation, 4 experts with at least 3 years' experience in programming and Computer Science Education to design the codebook, and also to rate the LLM-generated comments against the codebook criteria. 

\subsection{Codebook Development and First Round Evaluation}
To assess the overall quality of LLM-generated code comments, we developed a comprehensive codebook using criteria derived from \cite{fan2023exploring} based on eight essential criteria (Shown in Appendix). These criteria were carefully crafted to capture the key aspects of instructional quality in code comments, ensuring that the evaluation process is both rigorous and pedagogically relevant. This initial assessment provided a broad overview of the quality of the comments generated by different models. To gain deeper insights into the performance of the models, we conducted a statistical analysis of a 5-point Likert scale ratings, calculating the mean, standard deviation, and correlations among the criteria. Evaluations were conducted blind to ensure unbiased assessment.

\subsection{Second Round Expert Evaluation}
Building upon the findings from the first phase, we refined our evaluation approach in the second phase to include a set of binary criteria and a qualitative measure. The binary criteria were designed to be more focused and specific compared to the broader, more subjective Likert scale ratings. By concentrating on particular aspects of explanation and education, these criteria (see Appendix) allowed for a clearer differentiation between the capabilities of different models, providing a more nuanced understanding of their strengths and weaknesses in educational contexts.

To enhance the precision of our assessment, we removed the criterion ``Provides Enough Detail for Comprehension'' and introduced several new binary criteria that better captured the specific aspects of effective code comments. These criteria included ``Connects programming concepts to problem context,'' ``Consistent style and grammar,'' and ``Plain language,'' among others. Additionally, we introduced a qualitative measure, ``Functions as a friendly tutor,'' to evaluate the ability of the LLM-generated comments to engage and support learners in a manner similar to a human tutor.

The second round of expert evaluations was conducted using these refined criteria, with a blind rating process and both paired and global Likert ratings to ensure objectivity. A Chi-square analysis was performed on the binary criteria to gain further insights into the efficacy of LLM-generated code comments. While we refined the evaluation criteria after an initial review of responses to ensure relevance, we acknowledge the risk of overfitting. Nevertheless, this analysis allowed us to identify statistically significant differences in the performance of the models across various dimensions of explanation quality.

\begin{table}[h]
\centering
\begin{tabular}{l|c|c|c|c}
\toprule
Criterion & {GPT3.5} & {GPT4} & {Expert} & {Llama2} \\ 
\midrule
Explains Concepts \\ Used         & 3.85  & 4.03  & 3.63  & 2.83 \\
Identify Beginner \\ Mistakes     & 2.85  & 2.80  & 2.90  & 2.00 \\
Provides Code \\ Details                & 4.03  & 4.28  & 4.03  & 3.08 \\
Clarify Code \\ Behavior          & 3.90  & 3.75  & 4.13  & 2.85 \\
Break Down \\ Tasks               & 3.88  & 3.75  & 4.00  & 2.88 \\
Avoid Jargon                   & 4.30  & 4.08  & 4.18  & 2.98 \\
Explain Before \\ Code            & 3.70  & 3.50  & 3.85  & 2.73 \\
Simple Vocabulary              & 4.20  & 4.35  & 4.25  & 2.90 \\
\bottomrule
\end{tabular}
\caption{Comparison of the mean of model performance across various instructional quality criteria in code explanation tasks.}
\label{tab:stats_analysis}
\end{table}

\section{Analysis and Results}
\subsection{Statistical Analysis}
Our analysis utilized the mean values (shown in Table~\ref{tab:stats_analysis} associated with Kruskal-Wallis H test \cite{sprent2016applied} (See Appendix) to compare the performance of different language models (LLMs): GPT-4, GPT-3.5, Llama 2, and human experts. This test helped us identify significant differences across various evaluation criteria, providing insights into the strengths and weaknesses of each model in generating code comments. We specifically chose Kruskal-Wallis H test because our data is neither normal distributed nor parametric\cite{sprent2016applied}.

Clarity and Beginner-Friendliness: The results revealed that GPT-4 and human experts generally exhibited superior performance in clarity and beginner-friendliness. Both models scored consistently high across criteria such as avoiding jargon and simplifying vocabulary, making their comments more accessible to novices. Notably, GPT-4 outperformed Llama 2 in explaining concepts used, with significant p-values indicating its effectiveness in making complex ideas more comprehensible. Conversely, Llama 2 struggled in these areas, often failing to match the performance of its counterparts.

Concept Elucidation and Step-by-Step Guidance: In terms of concept elucidation, GPT-4 demonstrated an exceptional ability to clarify complex code behavior, closely followed by human experts. Both were proficient in breaking down tasks into manageable steps, facilitating better understanding for beginners. GPT-3.5, while slightly lagging behind GPT-4, still performed adequately and did not show significant statistical differences compared to the human expert, suggesting a strong capability in detailed explanation.

Comparative Performance Across Models: Llama 2’s performance was notably weaker, particularly in areas requiring detailed comments and the use of simple vocabulary. Statistical tests consistently showed Llama 2 as underperforming compared to GPT-3.5, GPT-4, and human experts, with low scores in breaking down tasks and clarifying code behavior, which are crucial for instructional quality.

These findings underscore the potential of advanced LLMs like GPT-4 to provide high-quality educational content that rivals and occasionally exceeds human expert comments. However, the variability among LLMs highlights the need for targeted improvements in models like Llama 2, especially in terms of beginner-friendliness and step-by-step guidance.

\begin{table*}[h]
\centering
\begin{tabular}{@{}lcrS[table-format=1.3]@{}}
\toprule
Comparison & Superior Model & {Chi-square Stat} & {P-value} \\
\midrule
Provide Inline Comments GPT35 vs GPT4 & GPT4 & 4.514 & 0.034 \\
Separate Code Chunks GPT35 vs GPT4 & GPT4 & 5.104 & 0.024 \\
High Level Overview GPT35 vs Human Expert & Human Expert & 7.656 & 0.006 \\
Discuss Complexity GPT4 vs Llama2 & GPT4 & 11.396 & 0.001 \\
Connect Concepts to Context GPT4 vs Llama2 & GPT4 & 6.416 & 0.011 \\
Explain Code Implementation GPT4 vs Llama2 & GPT4 & 8.182 & 0.004 \\
Logic Before Details Llama2 vs Human Expert & Human Expert & 5.161 & 0.023 \\
Discuss Complexity Llama2 vs Human Expert & Human Expert & 14.405 & 0.000 \\
Explain Concept Choice Llama2 vs Human Expert & Human Expert & 4.514 & 0.034 \\
\bottomrule
\end{tabular}
\caption{All significant results of Chi-Square Tests indicating superior model performance. The chi-square statistic measures the strength of evidence against the null hypothesis, with higher values indicating stronger evidence. The \( p \)-value indicates whether the observed differences are statistically significant, with values below 0.05 typically considered significant.}
\label{tab:significant_chi_square}
\end{table*}


\subsection{Chi Square Analysis}
The chi-square analysis was employed to assess the statistical significance of differences in performance between GPT-4 and other models, including Llama2, GPT-3.5, and human experts across various evaluation criteria. This section provides a detailed interpretation of the chi-square statistics, p-values, and degrees of freedom, which are consistently set at 1, reflecting the comparison between two groups per criterion. All the significant results of Chi-square analysis are shown in Table~\ref{tab:significant_chi_square}.

In the analysis of chi-square results across various comparative criteria, significant findings were observed that highlight differences in performance between GPT-4, Llama2, and GPT-3.5, as well as between these models and human experts. Among these, a notable outcome was the comparison between GPT-4 and Llama2 in discussing complexity, where GPT-4 demonstrated superior capability with a chi-square statistic of 11.40 and a p-value of 0.001. Similarly, GPT-4 showed a better ability to connect concepts to context than Llama2, indicated by a chi-square statistic of 6.42 and a p-value of 0.011.

No significant differences were observed between GPT-4 and human experts, suggesting that GPT-4’s performance is on par with human levels within the bounds of statistical significance. This parity is an indication of GPT-4's advanced capabilities in mimicking human reasoning and explanation, which are critical in educational and professional settings.
Conversely, GPT-3.5 demonstrated limitations compared to human experts, particularly in providing high-level overviews, where it lagged behind with a chi-square statistic of 7.66 and a p-value of 0.006. This suggests a potential area for improvement in GPT-3.5’s ability to synthesize and summarize information effectively.

Moreover, significant differences in discussing complexity were also observed between GPT-3.5 and Llama2, with GPT-3.5 underperforming (chi-square statistic of 5.10, p-value of 0.024), and between human experts and Llama2, where human experts excelled significantly (chi-square statistic of 14.40, p-value of 0.0001). These results underscore the variability in AI models' ability to handle complex analytical tasks compared to human experts, highlighting the necessity for ongoing enhancements in AI training methodologies to bridge these gaps.

\subsection{Friendliness Analysis}
Furthermore, we introduced the \textit{friendly\_tutor} criterion to qualitatively evaluate the LLM generated explanation's friendliness as a tutor, aiming to gather feedback from novice programming students. This criterion assesses how effectively these comments engage and support learners, fostering a supportive learning environment crucial for motivation and success in introductory computer science education. The ratings provided insights into the potential of language models to offer personalized support and enhance the learning experience for beginners.

The Mann-Whitney U test revealed significant differences in the perceived friendliness of the models' comments. GPT-4 outperformed both GPT-3.5 and Llama2 in friendliness, with U-statistics of 300.5 and 322.5, and p-values of 0.0017 and 0.0003, respectively. These results suggest that GPT-4's comments are perceived as significantly more friendly and supportive for beginners compared to those generated by GPT-3.5 and Llama2.
The human\_expert model demonstrated mixed results. It was perceived as more friendly than Llama2 (p=0.0023) but did not statistically differ from GPT-3.5 and GPT-4 in friendliness ratings. This indicates that while human experts are effective, the gap between human-generated and certain LLM-generated comments (specifically GPT-4) in perceived friendliness is narrowing.

The results indicate the importance of the \textit{friendly\_tutor} criterion in distinguishing how different models are perceived by learners. The findings suggest that GPT-4 excels in delivering comments that meet both the factual and emotional needs of novice programmers.

\section{Conclusion \& Discussion}
This study evaluated the instructional quality of code comments generated by Large Language Models (LLMs) compared to those developed by human experts, focusing on their application in novice programming education.

\textbf{RQ 1: Generation Quality}
Our analysis, utilizing criteria such as clarity, beginner-friendliness, and step-by-step guidance, indicates that comments from GPT-4 are often comparable to those developed by experts. Specifically, GPT-4 excels in clarity and beginner-friendliness, closely aligning with expert comments in terms of educational value.

\textbf{RQ 2: Strengths and Weaknesses}
The strengths of LLM-generated comments, particularly from GPT-4, include effective concept elucidation and comprehensive step-by-step guidance that facilitate understanding complex programming concepts. However, weaknesses were observed in models like Llama 2, which struggled with maintaining simplicity in vocabulary and providing sufficiently detailed comments for complete beginner comprehension.

\textbf{RQ 3: For Novice Programmers}
Considering criteria critical to novice programmers—such as explanation clarity, ability to identify and correct beginner mistakes, and the provision of detailed, jargon-free instructions—GPT-4 aligns more closely with the instructional needs of novice programmers than comments from other LLMs and closely matches the effectiveness of human experts. While GPT-4 shows strong performance in clarity and beginner-friendliness compared to GPT-3.5 and LLaMA-2, these findings are based on a limited dataset. Further research with a broader range of models and a larger dataset is needed to confirm its suitability for generating educational content for beginners.

In conclusion, while some LLMs like GPT-4 show potential to rival or even surpass human experts in certain aspects of code explanation, the performance variability among different LLMs underscores the need for ongoing improvements and customizations tailored to the specific educational contexts and needs of novice programmers. The findings of this study advocate for a nuanced application of LLMs in educational tools, ensuring they are leveraged in ways that genuinely enhance learning outcomes.

\section{Limitations}
It is important to acknowledge the limitations of this study, including the limited dataset, the subjectivity of expert evaluations, and the absence of user studies with novice programmers. Future research should focus on expanding the dataset, refining evaluation criteria, conducting user studies, exploring multimodal comments, and comparing LLMs with other AI-driven approaches to develop robust and effective tools for enhancing computer science education.

Despite these limitations, this study serves as a preliminary investigation into the potential of LLMs for generating code comments and lays the groundwork for future research in this area. By addressing the identified limitations and expanding the scope of the study, researchers can gain a more comprehensive understanding of how LLMs can be leveraged to support novice programmers in their learning journey, ultimately contributing to the development of effective AI-driven educational resources in computer science education.

\newpage
\bibliographystyle{abbrv}
\bibliography{sigproc}  

\begin{thebibliography}{10}

\bibitem{baierl2023applications}
J.~D. Baierl.
\newblock {\em Applications of Large Language Models in Education: Literature Review and Case Study}.
\newblock PhD thesis, University of California, Los Angeles, 2023.

\bibitem{Cui2022CodeExpEC}
H.~Cui, C.~Wang, J.~Huang, J.~P. Inala, T.~Mytkowicz, B.~Wang, J.~Gao, and N.~Duan.
\newblock Codeexp: Explanatory code document generation.
\newblock In {\em Conference on Empirical Methods in Natural Language Processing}, 2022.

\bibitem{cui2022codeexp}
H.~Cui, C.~Wang, J.~Huang, J.~P. Inala, T.~Mytkowicz, B.~Wang, J.~Gao, and N.~Duan.
\newblock Codeexp: Explanatory code document generation.
\newblock {\em arXiv preprint arXiv:2211.15395}, 2022.

\bibitem{fan2023exploring}
A.~Fan, H.~Zhang, L.~Paquette, and R.~Zhang.
\newblock Exploring the potential of large language models in generating code-tracing questions for introductory programming courses.
\newblock In {\em Findings of the Association for Computational Linguistics: EMNLP 2023}, pages 7406--7421, 2023.

\bibitem{leinonen2023comparing}
J.~Leinonen, P.~Denny, S.~MacNeil, S.~Sarsa, S.~Bernstein, J.~Kim, A.~Tran, and A.~Hellas.
\newblock Comparing code explanations created by students and large language models.
\newblock {\em arXiv preprint arXiv:2304.03938}, 2023.

\bibitem{leinonen2023using}
J.~Leinonen, A.~Hellas, S.~Sarsa, B.~Reeves, P.~Denny, J.~Prather, and B.~A. Becker.
\newblock Using large language models to enhance programming error messages.
\newblock In {\em Proceedings of the 54th ACM Technical Symposium on Computer Science Education V. 1}, pages 563--569, 2023.

\bibitem{li2023explaining}
J.~Li, S.~Tworkowski, Y.~Wu, and R.~Mooney.
\newblock Explaining competitive-level programming solutions using llms.
\newblock 2023.

\bibitem{macneil2022automatically}
S.~MacNeil, A.~Tran, J.~Leinonen, P.~Denny, J.~Kim, A.~Hellas, S.~Bernstein, and S.~Sarsa.
\newblock Automatically generating cs learning materials with large language models.
\newblock {\em arXiv preprint arXiv:2212.05113}, 2022.

\bibitem{macneil2022generating}
S.~MacNeil, A.~Tran, D.~Mogil, S.~Bernstein, E.~Ross, and Z.~Huang.
\newblock Generating diverse code explanations using the gpt-3 large language model.
\newblock In {\em Proceedings of the 2022 ACM Conference on International Computing Education Research-Volume 2}, pages 37--39, 2022.

\bibitem{Madaan2022LanguageMO}
A.~Madaan, S.~Zhou, U.~Alon, Y.~Yang, and G.~Neubig.
\newblock Language models of code are few-shot commonsense learners.
\newblock {\em ArXiv}, 2022.

\bibitem{madaan2022language}
A.~Madaan, S.~Zhou, U.~Alon, Y.~Yang, and G.~Neubig.
\newblock Language models of code are few-shot commonsense learners.
\newblock {\em arXiv preprint arXiv:2210.07128}, 2022.

\bibitem{meyer2023chatgpt}
J.~G. Meyer, R.~J. Urbanowicz, P.~C. Martin, K.~O’Connor, R.~Li, P.-C. Peng, T.~J. Bright, N.~Tatonetti, K.~J. Won, G.~Gonzalez-Hernandez, et~al.
\newblock Chatgpt and large language models in academia: opportunities and challenges.
\newblock {\em BioData Mining}, 16(1):20, 2023.

\bibitem{openai2023gpt4}
OpenAI.
\newblock Gpt-4 technical report, 2023.

\bibitem{sprent2016applied}
P.~Sprent and N.~C. Smeeton.
\newblock {\em Applied nonparametric statistical methods}.
\newblock CRC press, 2016.

\bibitem{touvron2023llama}
H.~Touvron, L.~Martin, K.~Stone, P.~Albert, A.~Almahairi, Y.~Babaei, N.~Bashlykov, S.~Batra, P.~Bhargava, S.~Bhosale, D.~Bikel, L.~Blecher, C.~C. Ferrer, M.~Chen, G.~Cucurull, D.~Esiobu, J.~Fernandes, J.~Fu, W.~Fu, B.~Fuller, C.~Gao, V.~Goswami, N.~Goyal, A.~Hartshorn, S.~Hosseini, R.~Hou, H.~Inan, M.~Kardas, V.~Kerkez, M.~Khabsa, I.~Kloumann, A.~Korenev, P.~S. Koura, M.-A. Lachaux, T.~Lavril, J.~Lee, D.~Liskovich, Y.~Lu, Y.~Mao, X.~Martinet, T.~Mihaylov, P.~Mishra, I.~Molybog, Y.~Nie, A.~Poulton, J.~Reizenstein, R.~Rungta, K.~Saladi, A.~Schelten, R.~Silva, E.~M. Smith, R.~Subramanian, X.~E. Tan, B.~Tang, R.~Taylor, A.~Williams, J.~X. Kuan, P.~Xu, Z.~Yan, I.~Zarov, Y.~Zhang, A.~Fan, M.~Kambadur, S.~Narang, A.~Rodriguez, R.~Stojnic, S.~Edunov, and T.~Scialom.
\newblock Llama 2: Open foundation and fine-tuned chat models, 2023.

\bibitem{xu2022systematic}
F.~F. Xu, U.~Alon, G.~Neubig, and V.~J. Hellendoorn.
\newblock A systematic evaluation of large language models of code.
\newblock In {\em Proceedings of the 6th ACM SIGPLAN International Symposium on Machine Programming}, pages 1--10, 2022.

\bibitem{yan2023practical}
L.~Yan, L.~Sha, L.~Zhao, Y.~Li, R.~Martinez-Maldonado, G.~Chen, X.~Li, Y.~Jin, and D.~Ga{\v{s}}evi{\'c}.
\newblock Practical and ethical challenges of large language models in education: A systematic literature review.
\newblock {\em arXiv preprint arXiv:2303.13379}, 2023.

\bibitem{ye2023comprehensive}
J.~Ye, X.~Chen, N.~Xu, C.~Zu, Z.~Shao, S.~Liu, Y.~Cui, Z.~Zhou, C.~Gong, Y.~Shen, J.~Zhou, S.~Chen, T.~Gui, Q.~Zhang, and X.~Huang.
\newblock A comprehensive capability analysis of gpt-3 and gpt-3.5 series models, 2023.

\end{thebibliography}
%
\newpage
\onecolumn
\appendix
This is the prompt to generate the code comments: ``Here is the prompt to generate code comments (with details omitted): Assume I am a novice programmer, am familiar with the problem statement and solution code, but need your help to fully understand the reasoning behind the solution. Write the markdown Java code with comments by following this procedure: 1. Given this code snippet, group lines into chunks of codes by functionality or other similar means. 2. First, describe the use, rationale, complexity, higher level approach of the solution and how it generally works in Java comments at the beginning of the code. 3. Then, include Java comments for important steps of the approach to explain what the code is doing in detail. 4. Lastly, add Java comments for anything else that's appropriate. Please analyze the given Java solution code and explain it in a detailed, structured way in a format of Markdown source code along with the original code. The comments may include:  time complexity, space complexity, brief problem summary, algorithm used in the solution, high level step-by-step approach explanation, more detailed explanation of the approach, proof of correctness. Write the explanation clearly and avoid ambiguity and try to cover the above as much as you could.  Note that you only need to create the Markdown source code and nothing else. Please use only "//" comments. Solution Code:  {code}.''

Here is a table with the Pair-wise Kruskal-Wallis H test for each pair of the criteria in the first round of the evaluation.
\begin{table*}[ht!]
\centering
\scriptsize
\begin{tabular}{@{}lcccccccccccccccc@{}}
\toprule
 & \multicolumn{4}{c}{\textbf{Avoid Jargon}} & \multicolumn{4}{c}{\textbf{Break Down Tasks}} & \multicolumn{4}{c}{\textbf{Clarify Code Behavior}} & \multicolumn{4}{c}{\textbf{Explain Before Code}} \\ 
\cmidrule(lr){2-5} \cmidrule(lr){6-9} \cmidrule(lr){10-13} \cmidrule(lr){14-17}
 & HE & GPT4 & L2 & GPT3.5 & HE & GPT4 & L2 & GPT3.5 & HE & GPT4 & L2 & GPT3.5 & HE & GPT4 & L2 & GPT3.5 \\ 
\midrule
HE & - & 0.55 & 0.00 & 0.58 & - & 0.51 & 0.01 & 0.53 & - & 0.32 & 0.00 & 0.10 & - & 0.19 & 0.01 & 0.34 \\
GPT4 & 0.55 & - & 0.01 & 0.23 & 0.51 & - & 0.04 & 0.94 & 0.32 & - & 0.02 & 0.98 & 0.19 & - & 0.10 & 0.49 \\
L2 & 0.00 & 0.01 & - & 0.00 & 0.01 & 0.04 & - & 0.02 & 0.00 & 0.02 & - & 0.01 & 0.01 & 0.10 & - & 0.04 \\
GPT3.5 & 0.58 & 0.23 & 0.00 & - & 0.53 & 0.94 & 0.02 & - & 0.10 & 0.98 & 0.01 & - & 0.34 & 0.49 & 0.04 & - \\
\midrule
 & \multicolumn{4}{c}{\textbf{Explains Concepts Used}} & \multicolumn{4}{c}{\textbf{Identify Beginner Mistakes}} & \multicolumn{4}{c}{\textbf{Provides Detail}} & \multicolumn{4}{c}{\textbf{Simple Vocabulary}} \\ 
\cmidrule(lr){2-5} \cmidrule(lr){6-9} \cmidrule(lr){10-13} \cmidrule(lr){14-17}
 & HE & GPT4 & L2 & GPT3.5 & HE & GPT4 & L2 & GPT3.5 & HE & GPT4 & L2 & GPT3.5 & HE & GPT4 & L2 & GPT3.5 \\ 
\midrule
HE & - & 0.01 & 0.19 & 0.12 & - & 0.60 & 0.00 & 0.85 & - & 0.11 & 0.08 & 0.99 & - & 0.47 & 0.00 & 0.91 \\
GPT4 & 0.01 & - & 0.00 & 0.17 & 0.60 & - & 0.01 & 0.68 & 0.11 & - & 0.00 & 0.06 & 0.47 & - & 0.00 & 0.44 \\
L2 & 0.19 & 0.00 & - & 0.02 & 0.00 & 0.01 & - & 0.00 & 0.08 & 0.00 & - & 0.08 & 0.00 & 0.00 & - & 0.00 \\
GPT3.5 & 0.12 & 0.17 & 0.02 & - & 0.85 & 0.68 & 0.00 & - & 0.99 & 0.06 & 0.08 & - & 0.91 & 0.44 & 0.00 & - \\
\bottomrule
\end{tabular}
\caption{Pair-wise Kruskal-Wallis H test to indicate the significance.}
\label{tab:kruskal}
\end{table*}

Here is the table with all criteria names and short descriptions.
\begin{table*}[!ht]
\scriptsize
\centering
\begin{tabular}{@{} >{\raggedright\arraybackslash}p{13cm} l @{}}
\toprule
\textbf{Description} & \textbf{Variable Name} \\
\midrule
\multicolumn{2}{c}{\textbf{First Round Evaluation Criteria}} \\
Explains programming concepts used & explains\_concepts\_used \\
Helps identify coding mistakes beginners often make & identify\_beginner\_mistakes \\
Provides enough detail for comprehension & provides\_detail \\
Makes relationships between code and behavior clearer & clarify\_code\_behavior \\
Breaks down complex tasks step-by-step & break\_down\_tasks \\
Avoids unnecessary technical jargon & avoid\_jargon \\
Explains concepts before giving code specifics & explain\_before\_code \\
Uses simple vocabulary appropriate for beginners & simple\_vocabulary \\
\midrule
\multicolumn{2}{c}{\textbf{Second Round Evaluation Criteria}} \\
Provides high-level overview first & high\_level\_overview \\
Explains logic or approach before line-by-line details & logic\_before\_details \\
Is there a discussion of time and space complexity & discuss\_complexity \\
Connects programming concepts to problem statement context & connect\_concepts\_to\_context \\
Warns about potential mistakes, misunderstandings or edge cases in implementing the approach & warn\_potential\_mistakes \\
Explain why certain concepts (e.g., data structures) are chosen and how they fit the algorithm & explain\_concept\_choice \\
Provide inline comments & provide\_inline\_comments \\
Comment styles and code conventions should be consistent throughout & consistent\_comments\_style \\
Explains how code implements intended behavior or show how to trace the program & explain\_code\_implementation \\
It separates code into logical chunks/steps and provides comments accordingly, in an easy-to-follow sequence & separate\_code\_chunks \\
Uses common words and simple sentences & use\_simple\_language \\
It works as a friendly tutor & friendly\_tutor \\
\bottomrule
\end{tabular}
\caption{Evaluation Criteria and Corresponding Variable Names}
\label{tab:criteria_variables}
\end{table*}

\balancecolumns
\end{document}